\documentclass[journal=pre,showpacs,showkeys,notitlepage,nofootinbib,twocolumn,floatfix]{revtex4-1}
\pdfpagewidth=\paperwidth
\pdfpageheight=\paperheight

\usepackage{gensymb}
\usepackage{enumitem}
\usepackage{graphicx}
\usepackage{verbatim}
\usepackage{color}
\usepackage{amsmath,amsfonts,amsbsy,amssymb}
\usepackage{upgreek}
\usepackage{natbib}

\def\micro{\mu}

\begin{document}

\author{Paul N. Patrone}
\email{paul.patrone@nist.gov}
\author{Anthony J. Kearsley}
\author{Erica L. Romsos}
\author{Peter M. Vallone}
\affiliation{National Institute of Standards and Technology}

\date{\today}
\title{Improving Baseline Subtraction for Increased Sensitivity of Quantitative PCR Measurements}

\begin{abstract}
Motivated by the current COVID-19 health-crisis, we examine the task of baseline subtraction for quantitative polymerase chain-reaction (qPCR) measurements.  In particular, we present an algorithm that leverages information obtained from non-template and/or DNA extraction-control experiments to remove systematic bias from amplification curves.  We recast this problem in terms of mathematical optimization, i.e.\ by  finding the amount of control signal that, when subtracted from an amplification curve, minimizes background noise.  We demonstrate that this approach can yield a decade improvement in sensitivity relative to standard approaches, especially for data exhibiting late-cycle amplification.  {\it Critically, this increased sensitivity and accuracy promises more effective screening of viral DNA and a reduction in the rate of false-negatives in diagnostic settings.}  
\end{abstract}


\maketitle


Quantitative polymerase chain-reaction (qPCR) measurements have had a long and successful history as a diagnostic tool in the medical community \cite{PCRReview}.  However, preliminary evidence suggests that false-negatives are on the order of 30\% for detection of the SARS-CoV-2 virus \cite{Error}.  Moreover, a recent study suggests that these incorrect results are associated with asymptomatic carriers and/or early stages of infection, when viral loads may be low \cite{lowdetect}.  Thus, methods that can better characterize and increase sensitivity of qPCR measurements will be critical for both understanding and controlling the current outbreak.

In this greater context, the task of baseline correction remains an overlooked problem that limits the sensitivity of qPCR.  Typical approaches rely on local, empirical fits of the amplification curves at low cycle numbers $n\le 15$, i.e.\ before the onset of exponential growth.  One assumes that this local behavior can be extrapolated to high cycle numbers to model the global background structure.  However, it is well known that empirically motivated extrapolation can introduce significant errors, especially when data are fit with polynomials \cite{Powell}.  Moreover, it is unreasonable to assume that baseline corrections so extracted will be valid for high-cycle numbers (e.g.\ $n\ge 35$).  In this region, exquisite accuracy is needed to detect low initial template concentrations, and poor baseline subtraction will confound interpretation of measurements.  Clearly, more robust approaches based on global properties of the data are needed to realize the full sensitivity of qPCR. 

To address this problem, we present a new baseline-subtraction strategy that avoids the use of empirical models by directly leveraging the behavior of appropriate control experiments.  The main idea of our approach is to postulate that the measurement signal is a linear combination of fluorescence due to DNA reporters and an unknown amount of  {\it extraction-blank} (EB) signal.  We directly estimate the contribution of the latter by determining the amount that, when subtracted from the measured signal, minimizes the background noise.  We show that this approach yields up to a decade improvement in sensitivity relative to approaches based on linear fits, especially for amplification curves that only show growth at late cycles.  Because such curves correspond to low initial DNA template numbers,  we speculate that our approach can be used to decrease the rate of false-negatives in qPCR-based diagnostic tools.


A notable aspect of our approach is that it makes few assumptions about the measurement process, thereby increasing robustness.  In particular, denote: (i) the measured fluorescence at the $n$th cycle by $f_n$; (ii) an appropriately measured ``baseline'' signal (see below for more details) by $b_n$; and the ``true'' or corrected signal corresponding to the number of DNA strands by $\tau_n$.  Our main assumption can be stated mathematically as
\begin{align}
f_n = \tau_n + a b_n + c \label{eq:model_equation}
\end{align}
where $a$ is an unknown coefficient expressing the amount of baseline signal polluting the true one, and $c$ is an unknown constant offset.  Physically, we interpret the latter as arising from dark currents and/or electronics offsets in the photodetector setup \cite{Affine1}.  We hypothesize that the contribution from $b_n$ corresponds to slight fluorescence of buffer components or impurities carried over from the DNA extraction process.  Thus, the coefficient $a$ corresponds to the unknown variability with which these components are unintentionally added to each well.

\begin{figure*}[ht]
\includegraphics[width=14cm]{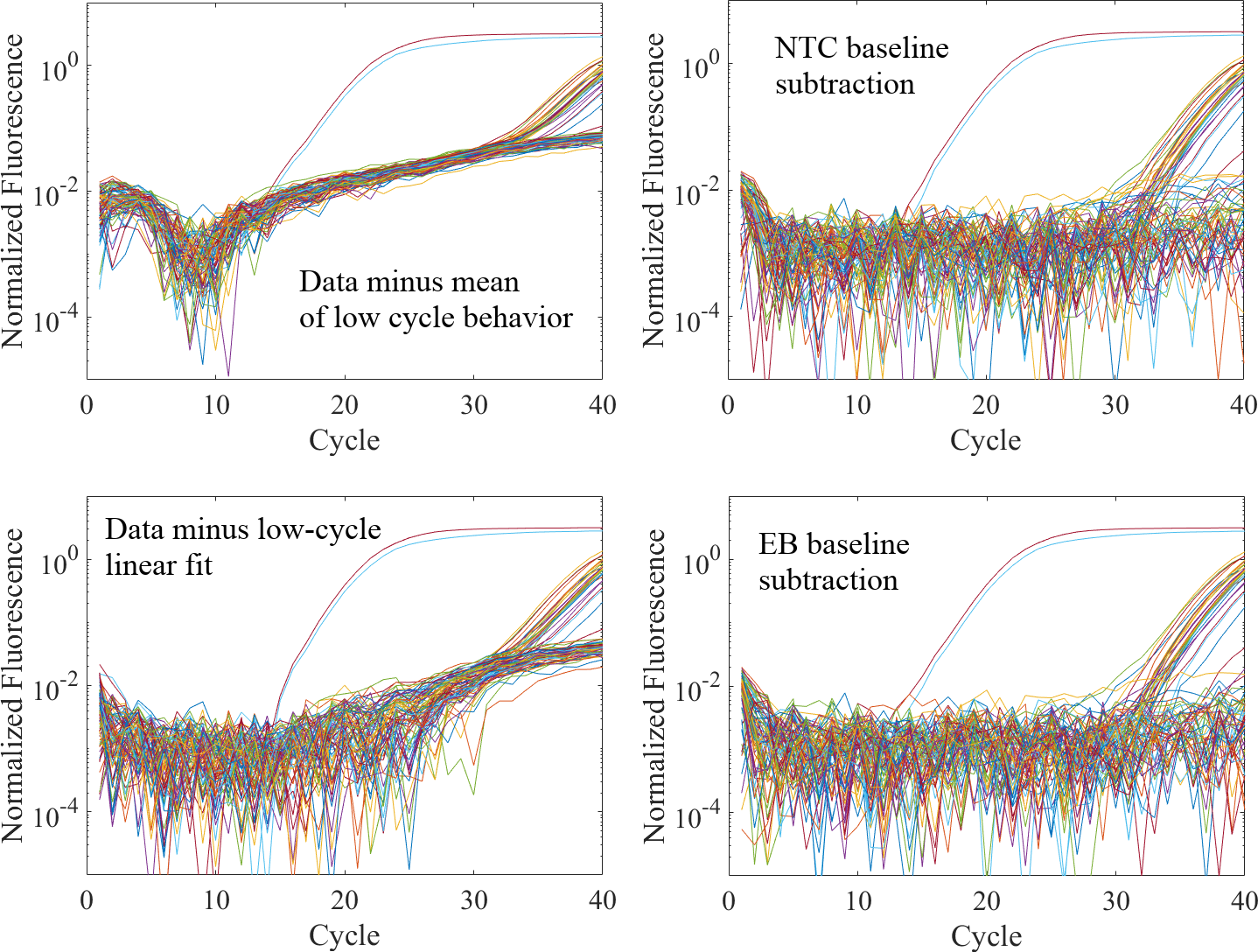}\caption{Baseline subtraction according to four different approaches for the ABY fluorescence channel on Plate 2; see the supplemental information for corresponding plots for the other data.  All plots show absolute value of the normalized fluorescence, so that no data is omitted on the log scale.  \textcolor{black}{In each plot, the two amplification curves exhibiting exponential growth around cycle 20 correspond to the Neat A and B components, which were used to set a scale for subsequent dilutions. } {\it Top left}: Raw data minus the mean values of cycles $3$ to $15$ computed individually for each curve.  Note that the systematic effects, which resemble square-root behavior, dominate the baseline at almost all cycles. {\it Bottom left}: Raw data minus a linear fit computed individually for each curve from cycles $3$ to $15$.  {\it Top Right:} Raw data corrected by optimizing of Eq.\ \eqref{eq:objective} using the mean NTC for $b_n$.  Note that the systematic effects present on the left plots have been eliminated.  {\it Bottom Right:} Raw data corrected by using the EB in place of the NTCs for $b_n$. Note that the systematic effects have been eliminated.}\label{fig:rawdat}
\end{figure*}

To determine $a$ and $c$, we first formulate an objective
\begin{align}
\mathcal L(a,c) &= \epsilon(a-1)^2+ \sum_{n=N_0}^{N_h} \frac{(f_n - ab_n - c)^2}{\Delta N -1} \nonumber \\
& \qquad + \left[\frac{1}{\Delta N}\sum_{n=N_0}^{N_h} f_n - ab_n - c \right]^2 \label{eq:objective}
\end{align} 
where $\epsilon$ is a regularization parameter satisfying $0<\epsilon\ll 1$, the term $\epsilon(a-1)^2$ penalizes large deviations of the baseline correction from magnitude of the measured control signal, and $\Delta N = N_h - N_0$.  Here $N_0$ and $N_h$ are lower and upper cycles for which $\tau_n$ is expected to be zero; below we discuss an adaptive routine by which these constants can be estimated.  Next, we minimize $\mathcal L(a,c)$ with respect to both variables, which determines them uniquely \cite{DS}.\footnote{More specifically, the objective is a strictly convex quadratic and thus possesses a unique minimizer.}  This optimization can be performed using a variety of numerical solvers in commercial software packages, e.g.\ the fminunc function in Matlab \cite{MATLAB}.\footnote{Certain commercial equipment, instruments, or materials are identified in this paper in order to specify the experimental procedure adequately. Such identification is not intended to imply recommendation or endorsement by the National Institute of Standards and Technology, nor is it intended to imply that the materials or equipment identified are necessarily the best available for the purpose.}  Finally, the corrected signal is estimated by re-writing Eq.\ \eqref{eq:model_equation} in terms of $\tau_n$ using the optimal values of $a$ and $c$.

\begin{figure*}
\includegraphics[width=14cm]{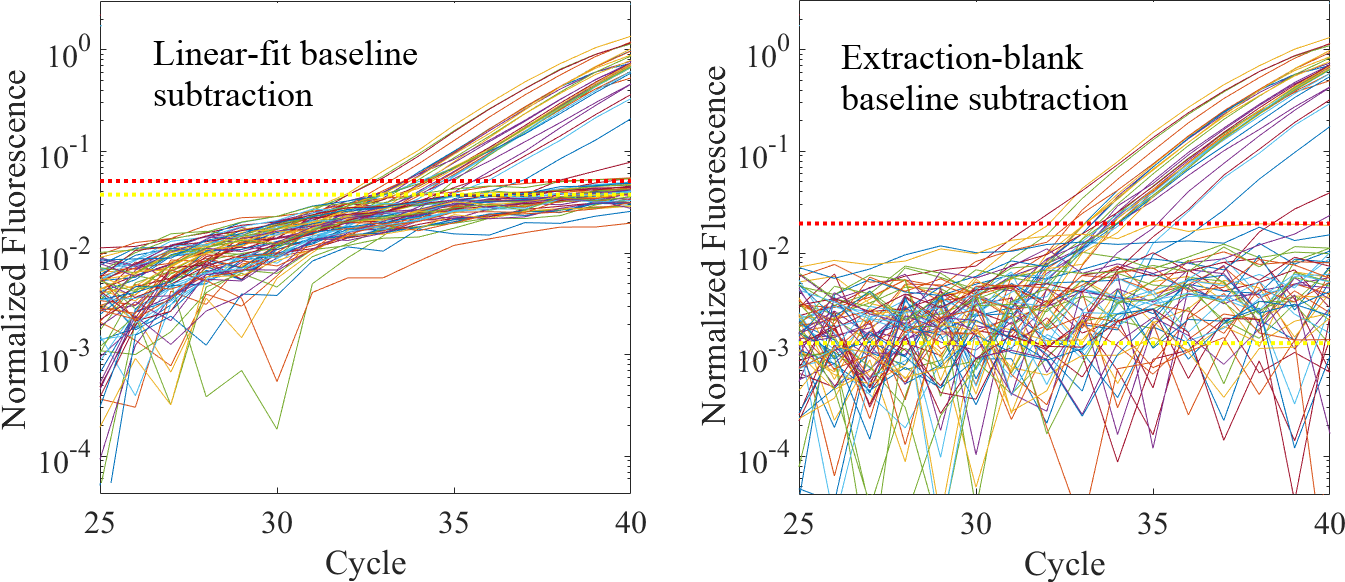}\caption{Close-up of the bottom subplots of Fig.\ \ref{fig:rawdat}.  In both sub-plots, the yellow dotted line is the mean value of all amplifications curves whose final value is less than 0.1.  The red dotted line is 2 standard deviations away from the mean.  Note that the extraction-blank baseline subtraction decreases the characteristic scales of the background by up to a decade.}\label{fig:comparison}
\end{figure*}

Several comments are in order.  The second and third terms appearing on the right-hand side (RHS) of Eq.\ \eqref{eq:objective} are the variance and squared-mean of $\tau_n$ across all cycles for which the fluorescence signal should be below the noise threshold.  Thus, in our approach, optimizing Eq.\ \eqref{eq:objective} ensures that $a$ and $c$ correspond to baseline subtractions that simultaneously suppress low-cycle noise and the corresponding mean.  While this is analogous to standard approaches that use a least-squares linear fit, the $b_n$ in Eq.\ \eqref{eq:model_equation} is measured empirically as opposed to postulated.  Moreover, as $b_n$ can be determined independently for each realization of a qPCR measurement (i.e.\ by running controls), variations arising from different machines, operators, etc.\ are inherently captured in the baseline subtraction process.

We also emphasize that the measurements generating $b_n$ should mimic as closely as possible the conditions of a sample containing DNA.  In our examples below, we find that baseline subtraction using EBs only provides marginal improvement  relative to using non-template controls (NTCs).  However, in setups where there is risk of contamination, EBs will provide a more accurate representation of the baseline because they are a product that represents the treatment of each unknown sample more effectively than NTCs, which are comprised of water or sterile buffer.  Thus, when possible, it is useful to run both to better characterize sources of the baseline.

To illustrate the usefulness of Eqs.\ \eqref{eq:model_equation} and \eqref{eq:objective} we performed a series of qPCR experiments using the Quantifiler Trio (Thermo Fisher) commercial qPCR chemistry.  

Extraction blanks were created by extracting six individual sterile cotton swabs (Puritain) using the Qiagen EZ1 Advanced XL and DNA Investigator kit (Qiagen).  290 $\micro$L of G2 buffer and 10 $\micro$L of Proteinase K were added to the tube and incubated in a thermal mixer (Eppendorf) at 56 \degree C for 15 minutes prior to being loaded onto the purification robot.  The ``Trace Tip Dance'' protocol was run on the EZ1 Advanced XL with elution of the DNA into 50 $\micro$L of TE (Qiagen).  After elution, all EBs were pooled into one tube for downstream analysis.  

Human DNA Quantitation Standard (Standard Reference Material 2372a) \cite{Standard} Component A and Component B were each diluted 10-fold.  Component A was diluted by adding 10 $\micro$L of DNA to 90 $\micro$L 10 mmol/L 2 amino 2 (hydroxymethyl) 1,3 propanediol hydrochloride (Tris HCl) and 0.1 mmol/L ethylenediaminetetraacetic acid disodium salt (disodium EDTA) using deionized water adjusted to pH 8.0 (TE${}^{-4}$, pH 8.0 buffer) from its certified concentration.  Component B was diluted by adding 8.65
$\micro$L of DNA to 91.35 $\micro$L of TE${}^{-4}$. From the initial 10-fold dilution, additional serial dilutions were performed down to 0.0024 pg into a regime to produce samples with high $C_q$ values ($>35$).   The concentration values targeted were determined empirically though the analysis of previous dilution data (not shown) and ranged from approximately 100 pg to 0.0024 pg, with an expectation that the 100 pg datapoint would fall between 28 and 29 $C_q$.  These two components were chosen due to Component A being male and Component B being female.  The qPCR chemistry used for these experiments is a chemistry which employs a Y-chromosomal target, which should amplify for Component A (male), but not for Component B (female), thereby permitting additional verification that the EBs are a reasonable proxy for baseline behavior.

For all qPCR reactions, Quantifiler Trio was used.  Each reaction consisted of 10 $\micro$L qPCR Reaction mix, 8 $\micro$L Primer mix, and 2 $\micro$L of sample (i.e. DNA, NTC, or EB) setup in a 96-well optical qPCR plate (Phoenix) and sealed with optical adhesive film (VWR).  After sealing the plate, it was briefly centrifuged to eliminate bubbles in the wells.  qPCR was performed on an Applied Biosystems 7500 HID instrument with the following 2-step thermal cycling protocol: 95 \degree C for 2 min followed by 40 cycle of 95 \degree C for 9 sec and 60 \degree C for 30 sec.  Data collection takes place at the 60 \degree C stage for 30 sec  for each of the cycles across all wells.  Upon completion of every run, data was analyzed in the HID Real Time qPCR Analysis Software v1.2 (Thermo Fisher) with a fluorescence threshold of 0.2.  Data was exported into an Excel file for further analysis.  

Two qPCR plates of 96 samples were run.  Plate one consisted of dilutions of Component A and B from 100 pg to 0.16 pg in triplicate, Neat Component A and B solutions in triplicate, 21 EBs and 21 NTCs, where NTC is TE-4 buffer without going through the extraction process.  Plate two consisted of dilutions of Component A and B from 0.31 pg to 0.002 pg in replicates of five, Neat Component A and B solutions in singlet, seven EBs, and seven NTCs.  From this data, we were able to generate curves with $C_q$ values ranging from approximately 27 to 40. 

For all custom data analyses using Eq.\ \eqref{eq:objective}, we set $N_0=3$ (to avoid transient effects associated with the first few cycles) and determined $N_h$ adaptively.  Specifically, we first performed an initial optimization of Eq.\ \eqref{eq:objective} by setting $N_h=15$ to obtain a rough estimate of the amplification curves.  We used these curves to estimate the cycle number $C_q$ at which any given fluorescence curve was closest to 0.1.  For $C_q\le 35$, we set $N_h=C_q-5$, which should put any given amplification curve in the noise regime.  For any value of $C_q>35$, we set $N_h=32$.  Then we reran the analysis to generate our final, baseline-subtracted amplification curves.  We used a value of $\epsilon=0.001$ for all calculations.\footnote{Data and analysis scripts are available upon request.}

Figure \ref{fig:rawdat} shows a representative plot comparing four baseline subtraction strategies for the ABY fluorescence channel on plate 2; see the supplemental information (SI) for other datasets.  For the top-left plot we subtracted the average of each curve from cycles 3 to 15; for the bottom-left, we subtracted a line fit to the same points.  The right plots use our new method, taking $b_n$ to be the NTC (top) or the EB (bottom).  Importantly, the two strategies that leverage average behavior of the data (left) introduce significant systematic effects with a characteristic scale of between 0.05 and 0.1 in terms of normalized fluorescence.  By contrast, baseline subtraction according to optimization of Eq.\ \eqref{eq:objective} (right) using either the NTCs or EBs eliminates all of the systematic effects and decreases the noise threshold by roughly a decade; see also Fig.\ \ref{fig:comparison}.

\begin{figure}
\includegraphics[width=7cm]{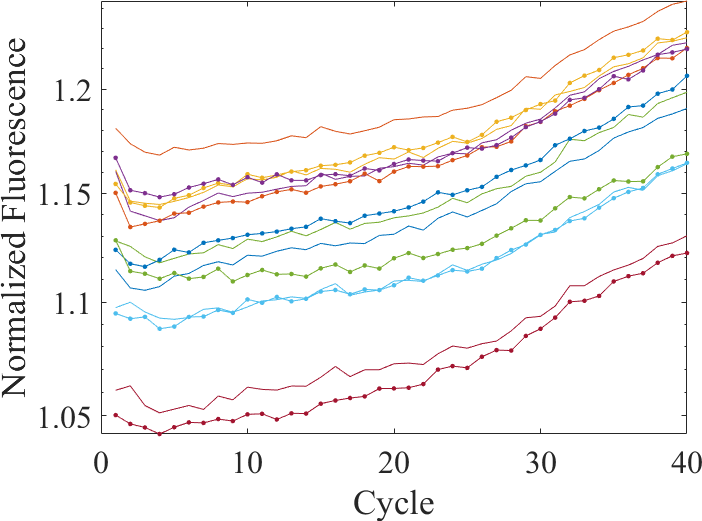}\caption{NTC (solid) and EB (solid with dots) curves measured with the data shown in Fig.\ \ref{fig:rawdat}.  Note that controls exhibit the same general behavior, which suggests their similar efficacy in removing baseline effects in the right plots of Figs.\ \ref{fig:rawdat}.}\label{fig:ntcsebs}
\end{figure}

Figure \ref{fig:ntcsebs} shows the corresponding NTC and EB measurements for this system.  From the figure, it is clear why the linear fit fails: neither the NTCs nor EBs can be modeled with lines past about cycle 20.  Moreover, a fit to the low-cycle data clearly underestimates the level of background.  Thus, subtracting a linear fit will underestimate the level of background, which explains the rise in data seen on the left plots of Fig.\ \ref{fig:rawdat}.  Moreover, the scale of the rise in the NTCs and EBs is consistent with the level of systematic effects seen in those approaches.

\begin{figure}
\includegraphics[width=7cm]{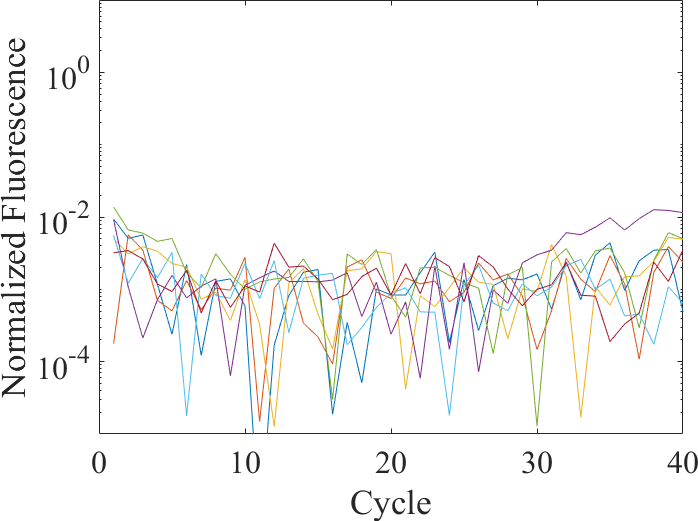}\caption{Baseline subtraction of the EBs using the mean EB according to optimization of Eq.\ \eqref{eq:objective}.  Note that the corrected curves remain below 0.01 and do not exhibit any significant systematic effects.}\label{fig:sanity}
\end{figure}

As a consistency check, we also applied our method to baseline subtraction of the EB curves.  Were this to generate false amplification, it would suggest lack of robustness in the method.  However, as Fig.\ \ref{fig:sanity} shows, background subtraction of EB (using the mean EB for $b_n$) does not generate any significant systematic effects, with all corrected datasets remaining within the noise.  


While more detailed studies are needed to characterize the improvements in absolute performance, the following conclusion is evident: {\it the sensitivity of qPCR measurements can be dramatically improved, essentially today, through our baseline-subtraction strategy leveraging NTC / EB signals; moreover, this can only improve the quality of any qPCR-based testing.}  In the long term, more formalized studies based on rigorous uncertainty quantification and consistency checks for fidelity of corrected data will be needed to realize the full potential of our approach; such work is currently on-going.  This will be important in developing a more complete epidemiological picture of viral spread.  In the short term, however, we anticipate that this improved sensitivity will allow for lower detection thresholds, thereby decreasing false-negatives in testing.

{\it Acknowledgements:} The authors thank Dr.\ Charles Romine for catalyzing a series of discussions that led to this work.  

{\it This work is a contribution of the National Institute of Standards and Technology and is not subject to copyright in the United States.}


\bibliography{PCR}

\begin{thebibliography}{8}%
\makeatletter
\providecommand \@ifxundefined [1]{%
 \@ifx{#1\undefined}
}%
\providecommand \@ifnum [1]{%
 \ifnum #1\expandafter \@firstoftwo
 \else \expandafter \@secondoftwo
 \fi
}%
\providecommand \@ifx [1]{%
 \ifx #1\expandafter \@firstoftwo
 \else \expandafter \@secondoftwo
 \fi
}%
\providecommand \natexlab [1]{#1}%
\providecommand \enquote  [1]{``#1''}%
\providecommand \bibnamefont  [1]{#1}%
\providecommand \bibfnamefont [1]{#1}%
\providecommand \citenamefont [1]{#1}%
\providecommand \href@noop [0]{\@secondoftwo}%
\providecommand \href [0]{\begingroup \@sanitize@url \@href}%
\providecommand \@href[1]{\@@startlink{#1}\@@href}%
\providecommand \@@href[1]{\endgroup#1\@@endlink}%
\providecommand \@sanitize@url [0]{\catcode `\\12\catcode `\$12\catcode
  `\&12\catcode `\#12\catcode `\^12\catcode `\_12\catcode `\%12\relax}%
\providecommand \@@startlink[1]{}%
\providecommand \@@endlink[0]{}%
\providecommand \url  [0]{\begingroup\@sanitize@url \@url }%
\providecommand \@url [1]{\endgroup\@href {#1}{\urlprefix }}%
\providecommand \urlprefix  [0]{URL }%
\providecommand \Eprint [0]{\href }%
\providecommand \doibase [0]{http://dx.doi.org/}%
\providecommand \selectlanguage [0]{\@gobble}%
\providecommand \bibinfo  [0]{\@secondoftwo}%
\providecommand \bibfield  [0]{\@secondoftwo}%
\providecommand \translation [1]{[#1]}%
\providecommand \BibitemOpen [0]{}%
\providecommand \bibitemStop [0]{}%
\providecommand \bibitemNoStop [0]{.\EOS\space}%
\providecommand \EOS [0]{\spacefactor3000\relax}%
\providecommand \BibitemShut  [1]{\csname bibitem#1\endcsname}%
\let\auto@bib@innerbib\@empty
\bibitem [{\citenamefont {Espy}\ \emph {et~al.}(2006)\citenamefont {Espy},
  \citenamefont {Uhl}, \citenamefont {Sloan}, \citenamefont {Buckwalter},
  \citenamefont {Jones}, \citenamefont {Vetter}, \citenamefont {Yao},
  \citenamefont {Wengenack}, \citenamefont {Rosenblatt}, \citenamefont
  {Cockerill},\ and\ \citenamefont {Smith}}]{PCRReview}%
  \BibitemOpen
  \bibfield  {author} {\bibinfo {author} {\bibfnamefont {M.~J.}\ \bibnamefont
  {Espy}}, \bibinfo {author} {\bibfnamefont {J.~R.}\ \bibnamefont {Uhl}},
  \bibinfo {author} {\bibfnamefont {L.~M.}\ \bibnamefont {Sloan}}, \bibinfo
  {author} {\bibfnamefont {S.~P.}\ \bibnamefont {Buckwalter}}, \bibinfo
  {author} {\bibfnamefont {M.~F.}\ \bibnamefont {Jones}}, \bibinfo {author}
  {\bibfnamefont {E.~A.}\ \bibnamefont {Vetter}}, \bibinfo {author}
  {\bibfnamefont {J.~D.~C.}\ \bibnamefont {Yao}}, \bibinfo {author}
  {\bibfnamefont {N.~L.}\ \bibnamefont {Wengenack}}, \bibinfo {author}
  {\bibfnamefont {J.~E.}\ \bibnamefont {Rosenblatt}}, \bibinfo {author}
  {\bibfnamefont {F.~R.}\ \bibnamefont {Cockerill}}, \ and\ \bibinfo {author}
  {\bibfnamefont {T.~F.}\ \bibnamefont {Smith}},\ }\href {\doibase
  10.1128/CMR.19.1.165-256.2006} {\bibfield  {journal} {\bibinfo  {journal}
  {Clinical Microbiology Reviews}\ }\textbf {\bibinfo {volume} {19}},\ \bibinfo
  {pages} {165} (\bibinfo {year} {2006})}\BibitemShut {NoStop}%
\bibitem [{\citenamefont {Ai}\ \emph {et~al.}(2020)\citenamefont {Ai},
  \citenamefont {Yang}, \citenamefont {Hou}, \citenamefont {Zhan},
  \citenamefont {Chen}, \citenamefont {Lv}, \citenamefont {Tao}, \citenamefont
  {Sun},\ and\ \citenamefont {Xia}}]{Error}%
  \BibitemOpen
  \bibfield  {author} {\bibinfo {author} {\bibfnamefont {T.}~\bibnamefont
  {Ai}}, \bibinfo {author} {\bibfnamefont {Z.}~\bibnamefont {Yang}}, \bibinfo
  {author} {\bibfnamefont {H.}~\bibnamefont {Hou}}, \bibinfo {author}
  {\bibfnamefont {C.}~\bibnamefont {Zhan}}, \bibinfo {author} {\bibfnamefont
  {C.}~\bibnamefont {Chen}}, \bibinfo {author} {\bibfnamefont {W.}~\bibnamefont
  {Lv}}, \bibinfo {author} {\bibfnamefont {Q.}~\bibnamefont {Tao}}, \bibinfo
  {author} {\bibfnamefont {Z.}~\bibnamefont {Sun}}, \ and\ \bibinfo {author}
  {\bibfnamefont {L.}~\bibnamefont {Xia}},\ }\href {\doibase
  10.1148/radiol.2020200642} {\bibfield  {journal} {\bibinfo  {journal}
  {Radiology}\ }\textbf {\bibinfo {volume} {0}},\ \bibinfo {pages} {200642}
  (\bibinfo {year} {2020})}\BibitemShut {NoStop}%
\bibitem [{\citenamefont {Liu}\ \emph {et~al.}(2020)\citenamefont {Liu},
  \citenamefont {Yan}, \citenamefont {Wan}, \citenamefont {Xiang},
  \citenamefont {Le}, \citenamefont {Liu}, \citenamefont {Peiris},
  \citenamefont {Poon},\ and\ \citenamefont {Zhang}}]{lowdetect}%
  \BibitemOpen
  \bibfield  {author} {\bibinfo {author} {\bibfnamefont {Y.}~\bibnamefont
  {Liu}}, \bibinfo {author} {\bibfnamefont {L.-M.}\ \bibnamefont {Yan}},
  \bibinfo {author} {\bibfnamefont {L.}~\bibnamefont {Wan}}, \bibinfo {author}
  {\bibfnamefont {T.-X.}\ \bibnamefont {Xiang}}, \bibinfo {author}
  {\bibfnamefont {A.}~\bibnamefont {Le}}, \bibinfo {author} {\bibfnamefont
  {J.-M.}\ \bibnamefont {Liu}}, \bibinfo {author} {\bibfnamefont
  {M.}~\bibnamefont {Peiris}}, \bibinfo {author} {\bibfnamefont {L.~L.~M.}\
  \bibnamefont {Poon}}, \ and\ \bibinfo {author} {\bibfnamefont
  {W.}~\bibnamefont {Zhang}},\ }\href {\doibase 10.1016/S1473-3099(20)30232-2}
  {\bibfield  {journal} {\bibinfo  {journal} {The Lancet Infectious Diseases}\
  } (\bibinfo {year} {2020}),\ 10.1016/S1473-3099(20)30232-2}\BibitemShut
  {NoStop}%
\bibitem [{\citenamefont {Powell}(1981)}]{Powell}%
  \BibitemOpen
  \bibfield  {author} {\bibinfo {author} {\bibfnamefont {M.}~\bibnamefont
  {Powell}},\ }\href {https://books.google.com/books?id=ODZ1OYR3w4cC} {\emph
  {\bibinfo {title} {Approximation Theory and Methods}}}\ (\bibinfo
  {publisher} {Cambridge University Press},\ \bibinfo {year}
  {1981})\BibitemShut {NoStop}%
\bibitem [{\citenamefont {Patrone}\ \emph {et~al.}()\citenamefont {Patrone},
  \citenamefont {Kearsley}, \citenamefont {Majikes},\ and\ \citenamefont
  {Liddle}}]{Affine1}%
  \BibitemOpen
  \bibfield  {author} {\bibinfo {author} {\bibfnamefont {P.}~\bibnamefont
  {Patrone}}, \bibinfo {author} {\bibfnamefont {A.}~\bibnamefont {Kearsley}},
  \bibinfo {author} {\bibfnamefont {J.}~\bibnamefont {Majikes}}, \ and\
  \bibinfo {author} {\bibfnamefont {J.~A.}\ \bibnamefont {Liddle}},\
  }\href@noop {} {\bibinfo  {journal} {Submitted}\ }\BibitemShut {NoStop}%
\bibitem [{\citenamefont {Dennis}\ and\ \citenamefont {Schnabel}(1996)}]{DS}%
  \BibitemOpen
\bibfield  {journal} {  }\bibfield  {author} {\bibinfo {author} {\bibfnamefont
  {J.~E.}\ \bibnamefont {Dennis}}\ and\ \bibinfo {author} {\bibfnamefont
  {R.~B.}\ \bibnamefont {Schnabel}},\ }\href {\doibase 10.1137/1.9781611971200}
  {\emph {\bibinfo {title} {Numerical Methods for Unconstrained Optimization
  and Nonlinear Equations}}}\ (\bibinfo  {publisher} {Society for Industrial
  and Applied Mathematics},\ \bibinfo {year} {1996})\BibitemShut {NoStop}%
\bibitem [{\citenamefont {MATLAB}(2013)}]{MATLAB}%
  \BibitemOpen
  \bibfield  {author} {\bibinfo {author} {\bibnamefont {MATLAB}},\ }\href@noop
  {} {\emph {\bibinfo {title} {version 8.1.0 (R2013a)}}}\ (\bibinfo
  {publisher} {The MathWorks Inc.},\ \bibinfo {address} {Natick,
  Massachusetts},\ \bibinfo {year} {2013})\BibitemShut {NoStop}%
\bibitem [{\citenamefont {Romsos}\ \emph {et~al.}(2018)\citenamefont {Romsos},
  \citenamefont {Kline}, \citenamefont {Duewer}, \citenamefont {Toman},\ and\
  \citenamefont {Farkas}}]{Standard}%
  \BibitemOpen
  \bibfield  {author} {\bibinfo {author} {\bibfnamefont {E.}~\bibnamefont
  {Romsos}}, \bibinfo {author} {\bibfnamefont {M.}~\bibnamefont {Kline}},
  \bibinfo {author} {\bibfnamefont {D.}~\bibnamefont {Duewer}}, \bibinfo
  {author} {\bibfnamefont {B.}~\bibnamefont {Toman}}, \ and\ \bibinfo {author}
  {\bibfnamefont {N.}~\bibnamefont {Farkas}},\ }\href {\doibase
  10.6028/NIST.SP.260-189} {\emph {\bibinfo {title} {Certification of Standard
  Reference Material 2372a Human DNA Quantitation Standard}}}\ (\bibinfo
  {publisher} {Natl. Inst. Stand. Technol. Spec. Publ. 260-189},\ \bibinfo
  {year} {2018})\BibitemShut {NoStop}%
\end{thebibliography}%

\end{document}